\documentclass{article}

\usepackage[english]{babel}
\usepackage[utf8]{inputenc}
\usepackage{textcomp}
\usepackage{microtype}
\usepackage{graphicx}
\usepackage{amsmath}
\usepackage{mathtools}
\usepackage{tikz} 
\usepackage{physics} 
\usetikzlibrary{positioning,arrows.meta,decorations.pathreplacing}

\usepackage[letterpaper,top=2cm,bottom=2cm,left=3cm,right=3cm,marginparwidth=1.75cm]{geometry}

\usepackage{amsmath}
\usepackage{graphicx}
\usepackage[
  colorlinks=true,
  allcolors=black,
  pdfborder={0 0 0},
  backref=page
]{hyperref}
\title{Bootstrapping Liquidity in BTC-Denominated Prediction Markets}
\author{Fedor Shabashev}

\begin{document}
\maketitle

\begin{abstract}
Prediction markets have gained adoption as on-chain mechanisms for aggregating information, with platforms such as Polymarket demonstrating demand for stablecoin-denominated markets. 
However, denominating in non-interest-bearing stablecoins introduces inefficiencies: participants face opportunity costs relative to the fiat risk-free rate, and Bitcoin holders in particular lose exposure to BTC appreciation when converting into stablecoins. 
This paper explores the case for prediction markets denominated in Bitcoin, treating BTC as a deflationary settlement asset analogous to gold under the classical gold standard. 

We analyse three methods of supplying liquidity to a newly created BTC-denominated prediction market: cross-market making against existing stablecoin venues, automated market making, and DeFi-based redirection of user trades. 
For each approach we evaluate execution mechanics, risks (slippage, exchange-rate risk, and liquidation risk), and capital efficiency. 
Our analysis shows that cross-market making provides the most user-friendly risk profile, though it requires active professional makers or platform-subsidised liquidity. 
DeFi redirection offers rapid bootstrapping and reuse of existing USDC liquidity, but exposes users to liquidation thresholds and exchange-rate volatility, reducing capital efficiency. 
Automated market making is simple to deploy but capital-inefficient and exposes liquidity providers to permanent loss. 

The results suggest that BTC-denominated prediction markets are feasible, but their success depends critically on the choice of liquidity provisioning mechanism and the trade-off between user safety and deployment convenience.
\end{abstract}

\section{Introduction}
Prediction markets have long been studied as mechanisms for aggregating information and forecasting uncertain events. 
Recent platforms such as Polymarket have demonstrated that on-chain prediction markets can attract substantial liquidity and user interest. 
However, most existing implementations are denominated in stablecoins such as USDC. 
While convenient for fiat-based accounting, stablecoin denomination introduces inefficiencies: bettors face an opportunity cost relative to the prevailing risk-free rate in fiat, and Bitcoin holders in particular must sacrifice BTC \cite{nakamoto2008bitcoin} exposure when converting into stablecoins. 

This paper explores the case for prediction markets denominated in deflationary assets such as Bitcoin. 
BTC denomination removes the fiat opportunity cost and allows BTC-native users to participate without losing exposure to Bitcoin’s long-term appreciation. 
At a conceptual level, this design mirrors hard-money frameworks such as the classical gold standard, where value is measured in a scarce settlement asset rather than an inflationary currency. 

We focus on the practical challenge of supplying liquidity to a newly created BTC-denominated prediction market. 
Three methods are examined in detail: (i) cross-market making against existing USDC venues, (ii) automatic market making, and (iii) DeFi-based redirection of user trades to external stablecoin markets. 
For each method we analyse execution mechanics, risk exposures (slippage, exchange-rate risk, liquidation risk), and capital efficiency. 
We argue that while DeFi redirection offers rapid bootstrapping, it imposes liquidation and FX risks on users, whereas cross-market making provides a safer user experience at the cost of higher operational demands on market makers.

The remainder of this paper examines how such markets can be bootstrapped with liquidity, and what trade-offs each method implies for users and platforms.

\subsection{Motivating Example: Long-Term Bets in Stablecoins}
In mid-2025, Polymarket introduced a 4.00\% annualized ``Holding Reward'' on selected long-dated political markets to encourage accurate long-term pricing, with rewards funded by the treasury and computed from hourly samples of users’ position value, paid daily \cite{polymarketHoldingRewards,gloria2025polymarket}. 

Without such a subsidy, bettors would face an opportunity cost relative to US treasuries yielding 4--5\% annually. 

Suppose in 2025, a bettor is 80\% confident in JD Vance winning the 2028 election. Rational pricing would require YES shares to trade below $64\%$ rather than $80\%$, since over four years the bettor sacrifices roughly 16\% by not holding treasuries. 

Similarly, a Bitcoin holder expecting 10\% annual appreciation faces a comparable opportunity cost when converting BTC to USDC: to justify a one-year bet, the expected edge must exceed 10\%. 

These examples illustrate how stablecoin denomination embeds fiat or BTC opportunity costs directly into market participation.

\section{Supplying liquidity}
Any prediction market is only useful when traders can buy and sell shares there. 
Just like equity and FX markets, prediction markets have professional market makers who submit limit orders and make money from bid-ask spread. 
However a newly created prediction market won't have market makers initially so it won't have liquidity. 
There are a few ways to bootstrap liquidity:
\begin{enumerate}
    \item Cross market making
    \item Automatic market making (AMM)
    \item Redirecting traders to other prediction markets, while acting as a custodian
\end{enumerate}
Let us discuss each of these methods in details:

\subsection{Cross market making}
Let's say Polymarket's orderbook has bid 0.50 USD and ask 0.51 USD for YES in a certain prediction market. 
Let's call this market the Source market.
Market maker can create buy and sell limit orders with the same prices (plus his commission) nominated in bitcoin, providing liquidity to the Downstream Market. 
So if Bitcoin price is 100,000 USD the market maker would set their buy limit order (bid) at $0.0000050 \text{ BTC}$ per share (equivalent to $\$0.50$) and their sell limit order (ask) at $0.0000051 \text{ BTC}$ per share (equivalent to $\$0.51$).

The market maker's goal is to provide liquidity by being willing to both buy and sell, profiting from the spread between the two prices. To do this, they convert the existing USD bid and ask prices into Bitcoin.

They match the market's bid price to buy from sellers.
    $$
    \frac{\$0.50 \text{ USD}}{100,000 \, \frac{\text{USD}}{\text{BTC}}} = 0.0000050 \text{ BTC}
    $$

They match the market's ask price to sell to buyers.
    $$
    \frac{\$0.51 \text{ USD}}{100,000 \, \frac{\text{USD}}{\text{BTC}}} = 0.0000051 \text{ BTC}
    $$

It is also important to reduce directional risk: market maker's exposure to one sided bets should be minimised. 

\begin{figure}[h!]
\centering
\resizebox{0.9\textwidth}{!}{%
\begin{tikzpicture}[>=Latex,thick]

\node[draw,minimum width=5cm,minimum height=3.5cm,label=below:{\Large Source Exchange}] (source) {};

\draw[red,very thick] ([yshift=0.8cm]source.west) -- ([yshift=0.8cm]source.east) node[left=5.2cm]{Ask};
\draw[green!70!black,very thick] ([yshift=-0.8cm]source.west) -- ([yshift=-0.8cm]source.east) node[left=5.2cm]{Bid};

\node[draw,minimum width=5cm,minimum height=3.5cm,right=7cm of source,label=below:{\Large Downstream Exchange}] (down) {};

\draw[red,very thick] ([yshift=1.2cm]down.west) -- ([yshift=1.2cm]down.east) node[left=5.2cm]{Ask};
\draw[green!70!black,very thick] ([yshift=-1.2cm]down.west) -- ([yshift=-1.2cm]down.east) node[left=5.2cm]{Bid};

\draw[dashed] ([yshift=0.8cm]down.west) -- ([yshift=0.8cm]down.east);
\draw[dashed] ([yshift=-0.8cm]down.west) -- ([yshift=-0.8cm]down.east);

\draw[<->] ([yshift=1.2cm]down.east) -- ([yshift=0.8cm]down.east)
  node[midway,right]{\(\Delta\)};

\draw[blue,->,thick] ([yshift=0.8cm]source.east) -- ([yshift=0.8cm]down.west);
\draw[blue,->,thick] ([yshift=-0.8cm]source.east) -- ([yshift=-0.8cm]down.west);

\end{tikzpicture}
}
\caption{Basic cross-market making. Orders on the downstream BTC-denominated exchange are mirrored from the source USD-denominated market. Dashed lines indicate the copied bid and ask, while solid lines represent the posted quotes in the downstream venue.}
\label{fig:crossmm_basic}
\end{figure}
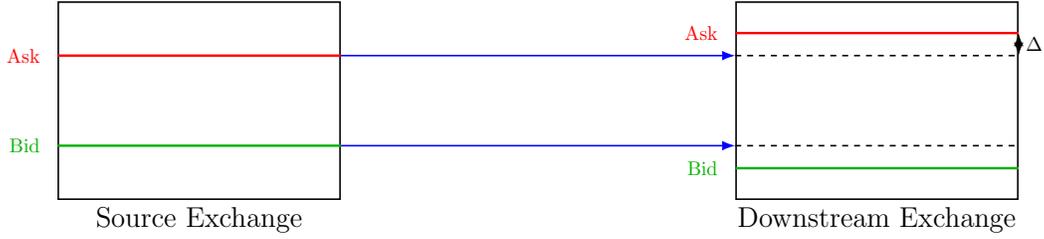

\begin{figure}[h!]
\centering
\resizebox{1\textwidth}{!}{%
\begin{tikzpicture}[>=Latex,thick]

\node[draw,minimum width=5cm,minimum height=3.5cm,label=below:{\Large Source Exchange}] (source) {};

\draw[red,very thick] ([yshift=0.8cm]source.west) -- ([yshift=0.8cm]source.east) node[left=5.2cm]{Ask \(\,= ask_s\)};
\draw[green!70!black,very thick] ([yshift=-0.8cm]source.west) -- ([yshift=-0.8cm]source.east) node[left=5.2cm]{Bid \(\,= bid_s\)};

\node[draw,minimum width=5cm,minimum height=3.5cm,right=7cm of source,label=below:{\Large Downstream Exchange}] (down) {};

\draw[red,very thick]   ([yshift=1.4cm]down.west) -- ([yshift=1.4cm]down.east) node[left=5.2cm]{Ask \(\,= ask_d\)};
\draw[green!70!black,very thick] ([yshift=-1.4cm]down.west) -- ([yshift=-1.4cm]down.east) node[left=5.2cm]{Bid \(\,= bid_d\)};

\draw[dashed] ([yshift=0.8cm]down.west) -- ([yshift=0.8cm]down.east);
\draw[dashed] ([yshift=-0.8cm]down.west) -- ([yshift=-0.8cm]down.east);

\draw[blue,->,thick] ([yshift=0.8cm]source.east) -- ([yshift=0.8cm]down.west);
\draw[blue,->,thick] ([yshift=-0.8cm]source.east) -- ([yshift=-0.8cm]down.west);

\draw[decorate,decoration={brace,amplitude=6pt}] ([xshift=0.28cm,yshift=0.8cm]down.east) -- ([xshift=0.28cm,yshift=1.1cm]down.east)
  node[midway,xshift=0.37cm] {\small fee};
\draw[decorate,decoration={brace,amplitude=6pt}] ([xshift=0.28cm,yshift=1.1cm]down.east) -- ([xshift=0.28cm,yshift=1.4cm]down.east)
  node[midway,xshift=0.9cm] {\small hedge cost};

\draw[decorate,decoration={brace,mirror,amplitude=6pt}] ([xshift=0.28cm,yshift=-0.8cm]down.east) -- ([xshift=0.28cm,yshift=-1.1cm]down.east)
  node[midway,xshift=0.37cm] {\small fee};
\draw[decorate,decoration={brace,mirror,amplitude=6pt}] ([xshift=0.28cm,yshift=-1.1cm]down.east) -- ([xshift=0.28cm,yshift=-1.4cm]down.east)
  node[midway,xshift=0.9cm] {\small hedge cost};

\node[align=left,anchor=west] at ([xshift=0.5cm,yshift=1.55cm]down.east)
  {\small \(ask_d = ask_s + fee + hedge\)};
\node[align=left,anchor=west] at ([xshift=0.5cm,yshift=-1.55cm]down.east)
  {\small \(bid_d = bid_s - fee - hedge\)};

\end{tikzpicture}
}
\caption{Downstream quotes widened/narrowed by fee and hedge cost relative to source quotes. Dashed = mirrored source; solid = final posted quotes.}
\label{fig:crossmm_fees}
\end{figure}
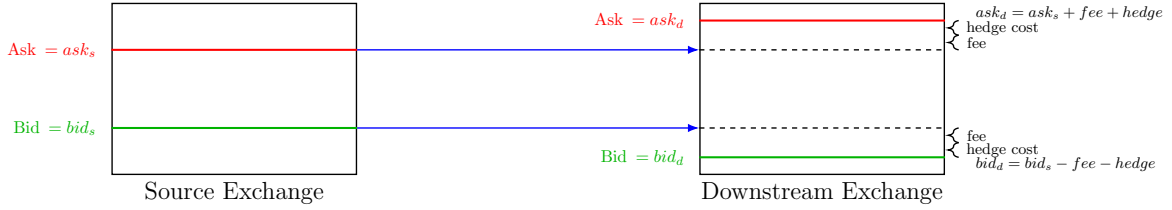

\paragraph{Cross exchange market making algorithm }
The algorithm for market making here goes as follows:
\begin{enumerate}
    \item Market maker quotes bid and ask prices as limit orders on the downstream market at the price of bid and ask on the source market plus commission: 
    $bid_d = bid_s - fee$ and $ask_d = ask_s + fee$, 
    where $bid_s$, $ask_s$ are the bid and ask on the source prediction market, and $bid_d$, $ask_d$ are the corresponding quotes on the downstream market.
    
    \item Once a downstream limit order is executed, the market maker immediately submits a market order on the source exchange on the opposite side of the market. 
    To illustrate this, assume both source and downstream markets are denominated in USDT. 
    This hedging step reduces exposure to directional risk. 
    Example: market maker quotes BUY limit orders for YES at $0.51$ USDT and for NO at $0.48$ USDT. 
    Suppose the NO order is filled (the maker buys NO at $0.48$).
    
    \item Once the market maker receives confirmation of the downstream execution, he immediately submits a market order on the source exchange to buy YES at $0.51$. 
    The order is executed, and now he holds one YES share in the source market and one NO share in the downstream market. 
    The combined cost is $0.51 + 0.48 = 0.99$ USDT for a position worth $1.00$ at resolution, capturing a $0.01$ profit before fees and slippage. 
    
    \item After these steps, the market maker’s position is:  
    Source market: 1 YES share.  
    Downstream market: 1 NO share.  
    Since YES + NO = 1 at resolution, the combined position is hedged. 
    To unwind, the market maker can now quote a SELL order for NO on the downstream market. 
    Once this is executed, he simultaneously sells the YES share on the source exchange. 
    In this way, the positions are closed on both markets, and the spread profit is realised. 
    
\end{enumerate}

\paragraph{Slippage risk.}
In practice, the hedge on the source may not execute exactly at the quoted price. 
    The order book may move in the time between downstream fill and source hedge, or the hedge size may consume multiple price levels. 
    If slippage exceeds the expected spread (e.g.\ the $0.01$ edge), the trade becomes unprofitable. 
    To mitigate this, the market maker must monitor liquidity depth, use protective limit orders, and quote only when the expected edge comfortably exceeds expected slippage and fees.

\paragraph{How to make it work with BTC (profitable example)}

To eliminate BTC/USD risk until resolution, the market maker buys a BTC \emph{put} expiring on the same date as the prediction market. 
Quotes on the BTC-denominated downstream venue must be widened to cover both trading fees and the put premium.

\subparagraph{Setup.}
Let $\text{BTCUSD} = 100{,}000$. 
On the source (USD) venue, suppose the best executable price for YES is $\$0.51$ (ask). 
On the downstream (BTC) venue, the market maker posts \emph{aggressive} buy interest for NO at
\[
\text{NO}_{d,\text{bid}} = 0.00000480\ \text{BTC} \quad (\approx \$0.48),
\]
i.e., a sufficiently low bid to create room for the hedge.
Assume the BTC put (expiry = market resolution, strike near spot) costs a premium of
\[
c_{\text{put}} = 0.000000020\ \text{BTC per share} \quad (\approx \$0.002 \text{ at } 100{,}000).
\]

\subparagraph{Trades.}
\begin{itemize}
  \item Buy 1 YES on the source at \$0.51.
  \item Buy 1 NO on the downstream at $0.00000480$ BTC ($\approx\$0.48$).
  \item Buy a BTC put expiring at resolution on notional $0.00000480$ BTC, premium $c_{\text{put}} \approx \$0.002$.
\end{itemize}

\subparagraph{Total cost (USD).}
\[
0.51\ (\text{YES})\ +\ 0.48\ (\text{NO in USD})\ +\ 0.002\ (\text{put})\ =\ 0.992\ <\ 1.00.
\]
This locks in a USD arbitrage of \$0.008 per complete set, \emph{before} trading fees on both venues. 
(The condition is $\text{YES}_{\text{ask}} + \text{NO}_{d,\text{bid}}\cdot P_{\text{BTCUSD}} + c_{\text{put}} < 1$.)

\subparagraph{Why the hedge works.}
At resolution, (YES + NO) = \$1 regardless of outcome. 
If BTCUSD falls, the BTC put compensates the USD loss on the BTC-denominated NO leg; if BTCUSD rises, the NO leg is worth more in USD and the put expires worthless—your initial edge already accounts for that premium. 
Thus USD P\&L is protected through to resolution.

\subparagraph{Quoting rule (include the put).}
When posting downstream prices in BTC, incorporate the put premium:
\[
\text{bid}_d = \frac{\text{bid}_s}{P_{\text{BTCUSD}}} - \varepsilon_{\text{fee}} - \frac{c_{\text{put}}}{P_{\text{BTCUSD}}},\qquad
\text{ask}_d = \frac{\text{ask}_s}{P_{\text{BTCUSD}}} + \varepsilon_{\text{fee}} + \frac{c_{\text{put}}}{P_{\text{BTCUSD}}},
\]
with tick-size rounding (round bids down, asks up). 
The spread must be wide enough that fills satisfying these quotes also satisfy the arbitrage inequality after all costs.

Figure~\ref{fig:crossmm_basic} illustrates the basic mirroring of source market quotes into the downstream venue. 
Figure~\ref{fig:crossmm_fees} extends this view by showing how fees and hedge costs are added to the downstream bid and ask, widening the spread relative to the mirrored quotes.

\subsection{Automatic market making}
Automatic market makers (AMMs) provide continuous liquidity without relying on active limit order submission by professional market makers. 
Instead of maintaining an order book, AMMs define a pricing rule that relates the relative price of assets to the quantities held in a liquidity pool. 
Traders interact directly with the pool, buying and selling at prices implied by this rule. 
In prediction markets, AMMs are attractive because they guarantee that shares can always be bought or sold, even in thin markets where order-book activity may be sporadic. 
Several AMM designs exist in the literature, most notably Hanson’s Logarithmic Market Scoring Rule (LMSR) \cite{hanson2007lmsr}, which guarantees bounded loss for the market maker but is less suitable for consumer-facing prediction markets. 

For this reason, modern platforms such as Manifold Markets use constant-product variants (CPMMs), which expose YES/NO shares with fixed \$1 payoffs and are easier for users to interpret. 
We now derive the CPMM pricing rule from first principles.

\subsubsection{Limitations of LMSR in practice}
While LMSR has elegant theoretical properties (bounded loss, continuous prices, and path independence), several features limit its practicality for consumer-facing prediction markets:
\begin{itemize}
  \item \textbf{Liquidity parameter $b$.} A single depth parameter $b$ governs both responsiveness and worst-case loss ($b\ln 2$ in a binary market). Too small: tiny trades cause large moves; too large: prices barely move and the operator underwrites a large subsidy. This makes scaling across heterogeneous, user-created markets difficult.
  \item \textbf{Asymptotic cost at the boundaries.} The dollars required to move price toward 0 or 1 grow logarithmically and diverge at the limits (e.g., $\Delta C = b\ln\!\frac{1-p_0}{1-p_1}$ for buys as $p_1\to 1$). Maintaining near-certainty prices or expressing extreme beliefs is capital-intensive.
  \item \textbf{User experience.} Traders conceptually “move the price” by paying a convex cost; the mapping to \$1-payout YES/NO shares is less direct than in constant-product designs, which many retail users find more intuitive.
\end{itemize}
These considerations help explain why modern platforms favor constant-product AMMs (CPMMs) that expose explicit YES/NO shares and inventory-based pricing, while LMSR remains valuable as a theoretical baseline and for curated markets \cite{hanson2007lmsr}.

\subsubsection{Derivation of CPMM from first principles}
Let's say we have a liquidity pool with two tokens: token $A$ and token $B$. \\
When someone wants to swap a certain amount of token $A$ and he wants to get some amount of token $B$ in return for his tokens $A$.\\
Let's assume that in the beginning the pool has $x$ amount of token $A$ and $y$ amount of token $B$. 
\begin{equation*}
\Delta x = x_{new} - x_{old} 
\end{equation*}
\begin{equation*}
\Delta y = y_{new} - y_{old}
\end{equation*}
Note that if $\Delta x > 0$ then $\Delta y < 0$ and vice versa.   

Now it is reasonable to expect that the exchange rate in the pool will be equal to: 
\begin{equation*}
P = \frac{y}{x}
\end{equation*}

Assuming $y = y(x)$
\begin{equation*}
 \lim_{\Delta x \to 0} \frac { \Delta y }{ \Delta x } =  \frac{- y}{x} 
\end{equation*} 

Or 
\begin{equation*}
    \frac{dy}{dx} = \frac{- y}{x}
\end{equation*}

Which is a first order linear differential equation and it is solvable by separating $y$ and $x$ and integrating each side.

\begin{equation*}
    \frac{dy}{dx} = \frac{- y}{x}
\end{equation*}
Separating the $x$ and $y$:
\begin{equation*}
    \frac{dy}{y} = \frac{-dx}{x}
\end{equation*}

Integrating each side:
\begin{equation*}
    \int \frac{dy}{y} = \int \frac{-dx}{x} 
\end{equation*}
After integrating we will get:

\begin{align}
\ln y   &= C_1 - \ln x\\
        &\Downarrow \\
y       &= \frac{C_2}{x}
\end{align}
From here we derive:
\begin{equation}
    xy = C
\end{equation}

We started with assumption that whenever a small amount of tokens is getting exchanged, the exchange rate closely follows the token amounts ratios. 
From that assumption we derived the constant product market making formula.

This yields the constant-product market-making formula. 

\subsubsection{CPMM in prediction markets}
In prediction markets, a CPMM is implemented as a pool containing YES and NO shares, where trades involve exchanging YES for NO shares and vice versa.
When a user wants to bet cash on YES, their cash is first converted into an equal bundle of YES and NO shares. Then, the NO shares are swapped for YES shares through the CPMM mechanism.
As a result, the user ends up holding only YES shares.

Among existing platforms, Manifold Markets is the most popular implementation of CPMM-based prediction markets by user count. 
Its design choices and liquidity trade-offs are described in public blog posts \cite{manifold2022marketmechanics}, and the full implementation is open-sourced \cite{manifoldGithub}.
For a broader systematization of AMM protocols and their variants, including their implications for slippage, capital efficiency, and liquidity provider risk, see Xu et al.\ \cite{xuamm}.

\subsubsection{Setting apriori price in CPMM}

A drawback of CPMMs in prediction markets is their lack of capital efficiency when it comes to subsidization. For example, suppose you want to subsidize a market with a starting probability of 33\% using \$100. Under a CPMM, you’d first convert your \$100 into 100 YES tokens and 100 NO tokens. To set up the pool, you’d then deposit 100 YES tokens and 50 NO tokens, since 50 / (100 + 50) = 1/3. (This may seem counterintuitive, but remember: in CPMMs, the more you buy of a token, the scarcer it becomes in the pool.) 
The leftover 50 NO tokens would remain in your portfolio as an implicit bet.

Rather than sticking with a standard CPMM, you can parameterize the pricing function around the initial probability:

\begin{equation}
    x^{p} y^{1-p} = C
\end{equation}

This way, the entire \$100 subsidy goes directly into the reserve pool. The result is greater liquidity for traders and no “spillover” position left in your portfolio.

This particular formula is implemented in Manifold Markets. 

\begin{figure}[h]
    \centering
    \includegraphics[width=0.7\linewidth]{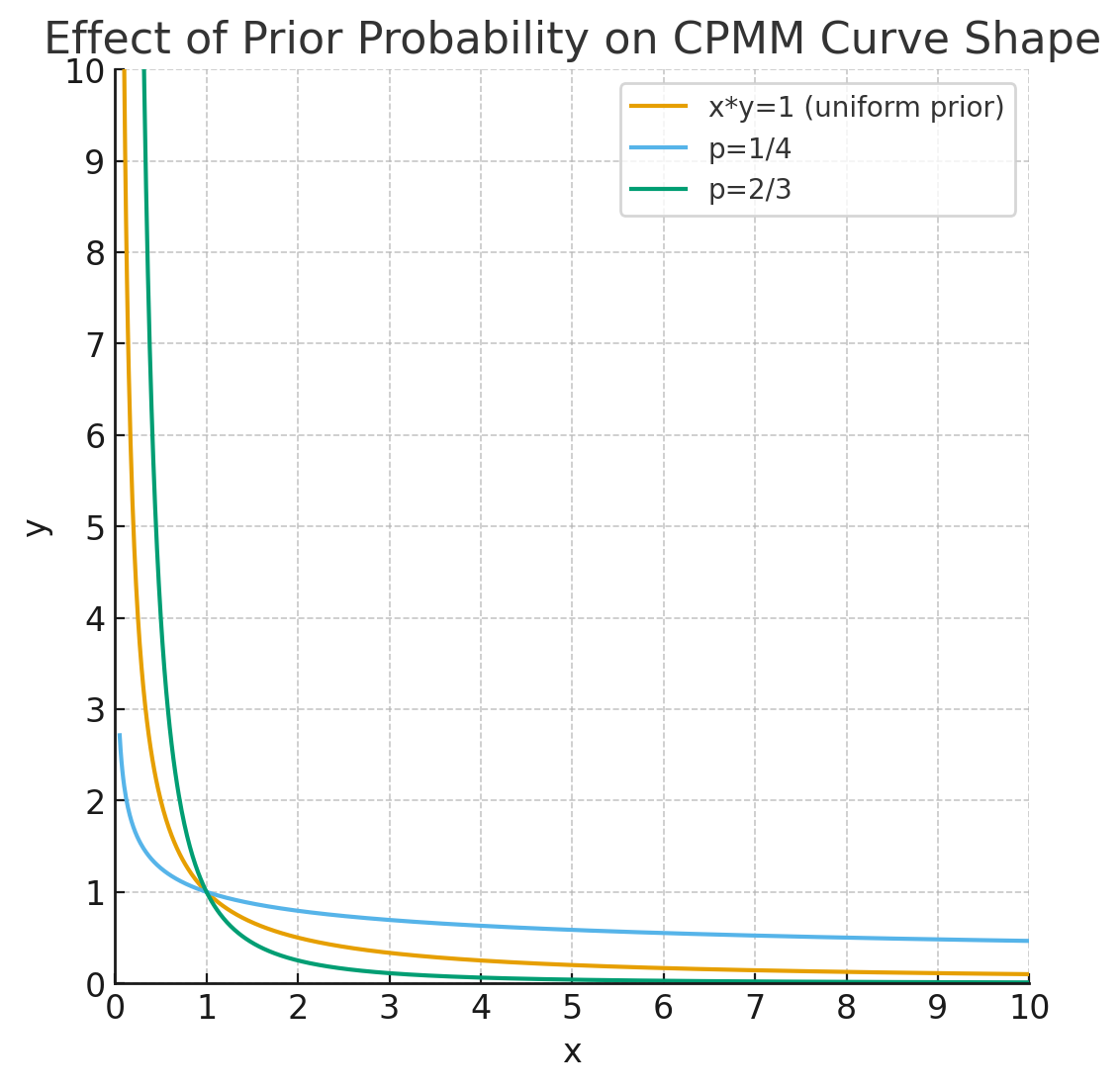}
    \caption{Effect of Prior Probability on CPMM Curve Shape}
    \label{fig:cpm_curve}
\end{figure}

\subsubsection{Permanent loss in CPMM}
Suppose a market maker seeds a CPMM pool by converting 100 USDT into 100 YES shares and 100 NO shares, with no fees collected.
As traders buy YES shares, the pool balances shift. Eventually, only 10 YES shares remain in the pool along with 1,000 NO shares.
This dynamic typically occurs when new external information makes the YES outcome more likely.

If the market then resolves as YES, the 10 YES shares are worth 10 USDT while the NO shares are worthless.
In other words, the pool’s value has fallen by 90\% compared to its original endowment.
For the pool creator, this is a poor outcome: their initial 100 USDT investment has effectively lost 90\% of its value.

\subsection{BTC payout via USDC infrastructure.}
An alternative implementation is to denominate contracts in BTC while routing execution through USDC-based venues. 
In this scheme, users deposit BTC, which is wrapped to wBTC on Polygon and used as collateral to borrow USDC on Aave \cite{aave2022v3}. 
The USDC is then deployed on Polymarket, and at resolution any USDC profit or loss is converted back into BTC for settlement with the user. 

This design has the advantage of reusing existing infrastructure (Aave, Polymarket) and avoids the need to build BTC-native trading rails. 
However, the payoff is not strictly BTC-nominated: since profits are realised in USDC and converted back at prevailing rates, the user’s BTC-denominated return depends on BTC/USD at settlement. 
From the perspective of a user who accounts wealth in BTC, this introduces foreign exchange risk. 

The benefit of this approach is practical implementability and access to deep liquidity; the drawback is that it only provides BTC accounting rather than a fixed BTC payoff per contract. 
A strict BTC-denominated design would require the platform itself to absorb or hedge BTC/USD risk, for example by running offsetting positions in BTC derivatives.

\paragraph{Borrowing versus selling BTC.}
A key advantage of the collateralised-borrowing approach is that the user retains exposure to the original BTC collateral. 
If BTC/USD rises substantially during the lifetime of the bet, selling BTC outright to obtain USDC would imply an opportunity cost, since the BTC would have been converted at a lower price. 
By contrast, borrowing USDC against BTC preserves the 1 BTC principal regardless of market movements. 

However, the incremental profit component is still realised in USDC and therefore subject to BTC/USD at conversion, but the base BTC position remains intact. 
This structure thus reduces the user’s risk exposure in BTC bull market scenarios, while still allowing profits to be paid in BTC terms.

\paragraph{Practical example with collateralised borrowing.}
In practice, collateralised lending platforms such as Aave do not allow 100\% loan-to-value ratios. 
For wBTC the maximum loan-to-value is currently around 72\%. 
This means that if a user deposits 1 BTC (worth \$115{,}910 at current prices), he can borrow up to about \$83{,}455 USDC against it. 

Suppose the user employs this USDC to purchase NO shares at \$0.80 on Polymarket. 
The number of shares purchased is
\[
\frac{83{,}455}{0.80} \approx 104{,}319 \ \text{NO shares}.
\]
If the event resolves NO, the shares pay out \$104{,}319, yielding a profit of 
\[
104{,}319 - 83{,}455 = 20{,}864 \ \text{USDC}.
\]

Converting this profit back into BTC at the entry price of 115{,}910 USD/BTC gives
\[
\frac{20{,}864}{115{,}910} \approx 0.180 \ \text{BTC}.
\]
The user therefore receives back his original 1 BTC collateral plus an additional 0.18 BTC, for a total of about 1.18 BTC. 

If BTC/USD rises before conversion, the USDC profit translates into fewer BTC; if BTC/USD falls, the profit translates into more BTC. 
Thus the 1 BTC collateral is always returned, but the incremental profit component remains USD-denominated until conversion.

\paragraph{Liquidation risk.}
The above calculation assumes that the 1 BTC collateral is always returned to the user, but this holds only in the absence of liquidation. 
On Aave, liquidation is triggered when the collateral-to-debt ratio falls below the liquidation threshold, which is currently around 82\% for wBTC. 
If a user borrows at the maximum 72\% LTV, even a 12--13\% decline in the BTC/USD price is sufficient to reach this threshold and cause liquidation. 
In such a case, part of the BTC collateral is sold at a discount to repay the debt, and the user receives back less than 1 BTC. 

For this reason, prudent users typically borrow at much lower LTV ratios (e.g.\ 30--40\%) to provide a safety buffer against BTC volatility. 
Thus, while collateralised borrowing preserves BTC exposure in normal conditions, it introduces the risk of losing a portion of the BTC principal if the market moves adversely.

\paragraph{Non-custodial redirection algorithm (BTC \(\rightarrow\) USDC \(\rightarrow\) BTC).}
The following on-chain flow routes BTC-denominated user orders to a USDC venue (e.g.\ Polymarket) without custody:

\begin{enumerate}
  \item \textbf{Deposit.} User transfers BTC to a bridge/wrapper; the protocol mints wBTC to a position-specific smart account (e.g.\ ERC-4337 or proxy safe) controlled by the user.
  \item \textbf{Collateralise.} Smart account supplies wBTC to Aave as collateral. Target a conservative LTV \(L^* < LTV_{\max}\) (e.g.\ \(L^*=0.4\)–\(0.5\)) and record liquidation threshold \(\Theta\).
  \item \textbf{Borrow.} Borrow USDC up to \(B = L^* \cdot V_{\mathrm{wBTC}}\) (in USD terms). Store initial health factor \(HF_0\).
  \item \textbf{Execute trade.} Contract routes USDC to the prediction venue (or a permissioned relayer) to buy the user’s shares (YES/NO). Trade receipts are held by the smart account (or a venue-specific escrow under the account’s control).
  \item \textbf{Risk loop (non-custodial).} A keeper/automation monitors:
    \begin{itemize}
      \item \textit{Health factor:} If \(HF \leq HF_{\mathrm{guard}}\) (e.g.\ 1.2), then \emph{auto-deleverage}: (i) sell a portion of the venue position back to USDC; (ii) repay USDC debt; (iii) optionally withdraw excess wBTC to restore \(HF\).
      \item \textit{LTV drift:} If \(LTV \uparrow\) due to BTC/USD fall, contract can (i) request user top-up of wBTC, or (ii) reduce borrow by partial repayment from any available USDC (fees, rebates, partial closes).
    \end{itemize}
  \item \textbf{Resolution or close.} Upon market resolution (or early close):
    \begin{enumerate}
      \item \emph{Settle:} Redeem USDC proceeds (may be zero if the bet lost) to the smart account.
      \item \emph{Repay:} Use all USDC to repay Aave principal + interest.
      \item \emph{Shortfall handling:} If proceeds \(<\) debt, the contract prompts the user to add USDC (or authorise a minimal wBTC swap \(\rightarrow\) USDC) to avoid liquidation and fully repay.
      \item \emph{Withdraw:} After debt is zero, withdraw wBTC collateral back to the smart account; unwrap/bridge to BTC and return to the user.
      \item \emph{Profit conversion:} Any surplus USDC (after full repayment) is swapped to wBTC and bridged to BTC for BTC-denominated payout.
    \end{enumerate}
  \item \textbf{Fail-safes.} If user is unresponsive and \(HF \to 1\): (i) auto-partial close of venue position; (ii) auto-repay from recovered USDC; (iii) as last resort, perform a minimal wBTC\(\rightarrow\)USDC swap to restore \(HF>1\). These steps are strictly bounded by on-chain caps the user set at deposit time.
\end{enumerate}

\noindent\textit{Parameters (suggested):} \(L^*=40\%\text{--}50\% \ll LTV_{\max}\), guard band \(HF_{\mathrm{guard}}\approx 1.2\), slippage caps on all swaps, and per-action gas/amount limits. The user retains ownership of the smart account; the protocol never takes custody. Repayment at resolution minimises interest accrual and liquidation risk; if the bet loses, the user can repay from external funds to reclaim full BTC collateral.

\subsection{Comparison of liquidity bootstrapping methods}

Cross market making and non-custodial DeFi redirection both allow users to place bets in BTC and receive payouts in BTC, but the mechanics and risk profiles differ substantially.

\paragraph{Cross market making.}
In this model, a professional market maker quotes bids and asks on the BTC-denominated downstream market while hedging on the source venue. 
The market maker absorbs all execution risks, including slippage, currency exposure, and inventory management. 
For users, the interface is simple and safe: they face no liquidation mechanics, and their BTC principal is only at risk through the outcome of the prediction market. 
The drawback is that the downstream venue depends on the presence of active market makers willing to provide liquidity. 
In practice, this dependency can be mitigated in the early stages by the platform itself acting as a subsidising market maker, until sufficient user activity attracts independent professional makers.

\paragraph{DeFi redirection.}
In the redirection model, user BTC is wrapped to wBTC, supplied as collateral to Aave, and used to borrow USDC that is then traded on Polymarket. 
Profits are converted back into BTC at resolution. 
This approach eliminates custodial risk and directly reuses Polymarket’s liquidity, but it introduces loan-to-value and liquidation thresholds. 
If BTC/USD falls sufficiently, liquidation can occur and part of the user’s BTC collateral may be lost, regardless of the market outcome. 
Borrowing at the maximum 72\% LTV implies liquidation after only a 12--13\% fall in BTC/USD, so in practice users must borrow at lower ratios (e.g.\ 30--40\%) to provide a buffer, which reduces capital efficiency. 
In addition, profits are realised in USDC and converted back at the prevailing BTC/USD rate, exposing users to exchange-rate risk on top of prediction market risk. 
Thus DeFi redirection trades off both capital efficiency and stability of BTC-denominated returns in exchange for faster deployment and reuse of existing USDC infrastructure.

\paragraph{Comparison.}
From a user perspective, cross market making offers greater safety since liquidation risk is eliminated, but it requires professional makers to actively operate. 
DeFi redirection is easier to bootstrap and scales with available DeFi infrastructure, but exposes users to collateral risks that are independent of prediction market outcomes. 
Automatic market making (AMM), while simple to deploy, suffers from permanent loss risks for liquidity providers and less efficient spreads for traders. 
These differences reflect a broader pattern observed in DeFi protocol design: each mechanism allocates risk and capital efficiency differently between users, liquidity providers, and market makers \cite{xu_defi_business_model}. 
Overall, cross market making provides the most user-friendly risk profile, whereas DeFi redirection trades user safety for faster deployment and access to external liquidity.

\section{Conclusion}
This paper considered prediction markets denominated in Bitcoin and examined practical methods for bootstrapping liquidity. 
Three approaches were analyzed: cross-exchange market making, automated market making via CPMMs, and DeFi-based redirection of user trades. 
Each method presents distinct trade-offs in terms of user risk, capital efficiency, and operational demands. 

Cross-exchange market making shields users from liquidation and currency risks, but requires professional makers or platform-subsidised liquidity to operate effectively. 
Automated market making ensures continuous liquidity and simple deployment, but is capital-inefficient and can result in permanent loss for liquidity providers. 
DeFi redirection leverages existing USDC infrastructure for rapid deployment, but shifts collateral and exchange-rate risks to users. 

Overall, the analysis shows how different liquidity models allocate risks and costs differently between users, liquidity providers, and market makers.

\bibliographystyle{alpha}

\end{document}